\documentclass[preprint,authoryear,12pt]{elsarticle-10052885v7}  

\usepackage{epsfig}

\usepackage{amssymb}
\usepackage{amsmath} 

\newcommand{\version}{v7}     
\renewcommand{\today}{arXiv:1005.2885 (\version)} 
\newcommand{\beq}{\begin{equation}}
\newcommand{\eeq}{\end{equation}}
\newcommand{\beqa}{\begin{eqnarray}}
\newcommand{\eeqa}{\end{eqnarray}}
\newcommand{\bsubeqs}{\begin{subequations}}
\newcommand{\esubeqs}{\end{subequations}}
\usepackage[ps2pdf,%
a4paper=true,%
breaklinks=true,%
colorlinks=true,%
pdfauthor={First Author et al.},%
pdftitle={Template for manuscripts in Advances in Space Research}%
]{hyperref}

\journal{Advances in Space Research 49 (2012) 213}  

\begin{document}

\begin{frontmatter}



\title{QCD-scale modified-gravity universe
}


\author{F.R. Klinkhamer
}
\address{Institute for the Physics and Mathematics of the universe,\\
    University of Tokyo, Kashiwa 277--8583, Japan\\
    and\\
    Institute for Theoretical Physics, University of Karlsruhe,\\
    Karlsruhe Institute of Technology, 76128 Karlsruhe,
    Germany$^{\,1}$  
    }
\fntext[footnote2]{Permanent address.}
\ead{frans.klinkhamer@kit.edu}




\begin{abstract}
A possible gluon-condensate-induced modified-gravity model with
$f(R) \propto |R|^{1/2}$ has been suggested previously. Here, a simplified
version is presented using the constant flat-spacetime equilibrium value of
the QCD gluon condensate and a single pressureless matter component (cold
dark matter, CDM). The resulting dynamical equations of a spatially-flat
and homogeneous Robertson--Walker universe are solved numerically.
This simple empirical model allows, in fact, for a careful treatment of the
boundary conditions and does not require a further scaling analysis as the
original model did. Reliable predictions are obtained for several observable
quantities of the homogeneous model universe. In addition, the estimator
$E_{G}$, proposed by Zhang et al. to search for deviations from standard
Einstein gravity, is calculated for linear sub-horizon matter-density
perturbations. The QCD-scale modified-gravity prediction for $E_{G}(z)$
differs from that of the $\Lambda\text{CDM}$ model by about $\pm 10\%$
depending on the redshift $z$.
\end{abstract}


\end{frontmatter}

\parindent=0.5 cm

\newpage
\setcounter{equation}{0}
\renewcommand{\theequation}{\arabic{section}.\arabic{equation}}
\section{Introduction}\label{sec:introduction}

It has been suggested
that the gluon condensate~\citep{ShifmanVainshteinZakharov1979,Narison1996}
of quantum chromodynamics [QCD] in the $q$--theory
framework~\citep{KV2008-statics,KV2008-dynamics,KV2010-CCP1}
may lead to a
particular type of $f(R)$ modified-gravity model~\citep{KV2009-gluoncond}.
The cosmology of this model was explored in \citet{K2010-gluoncondcosm}.
Here, the model is simplified even further, allowing for a detailed
study of the corresponding cosmological equations and eliminating
all arbitrariness.

The model of this article is, however, considered
from a different point of view than the one used
in the original article~\citep{KV2009-gluoncond}.
Here, the aim is not to derive the particular model used
from the underlying theory (general relativity and QCD),
but to consider the model \emph{per se} and to test
the hypothesis whether or not this model gives a reasonably
accurate description of the observed ``accelerating universe''
\citep[cf.][and references therein]{Riess-etal1998,Perlmutter-etal1999,
Komatsu-etal2009}.
In other words, the approach is in the sprit of Kepler
rather than Newton.

Still, the model is not chosen at random from all possible
$f(R)$ modified-gravity models, but obeys certain
physically motivated selection criteria, as discussed in
Section~\ref{subsec:Selection-criteria}.
The selected QCD-scale modified-gravity model is detailed
in Section~\ref{subsec:Representative-model},
with a few technical points relegated to
Section~\ref{subsec:Additional-remarks}.

The corresponding cosmological model with a single matter
component (cold dark matter, CDM) is introduced in
Section~\ref{subsec:From-three-components-to-one}.
The resulting differential equations are given
in Section~\ref{subsec:Dimensionless-equations}.
Three observable quantities of the homogeneous
cosmological model are then discussed in
Section~\ref{subsec:Observables-quantities}.
A fourth observable, this time of the inhomogeneous
cosmological model, is also considered, namely,
the estimator $E_{G}$, which was designed~\citep{Zhang-etal2007} to
characterize the linear growth of matter-density perturbations and which
has been measured recently for moderate redshifts~\citep{Reyes-etal2010}.
In Section~\ref{sec:Benchmark-LambdaCDM-results},
the same four observable quantities are calculated analytically
for the standard $\Lambda\text{CDM}$--model
universe~\citep{Weinberg1972,Carroll-etal1992,SahniStarobinsky2000,
PeeblesRatra2003,Perivolaropoulos2010}
and may serve as benchmark results.

The numerical results for both homogeneous and inhomogeneous
observable quantities of the QCD-scale modified-gravity
universe are discussed in Section~\ref{sec:Numerical-results}
and compared with those of the $\Lambda\text{CDM}$ universe.
Final comments are presented in Section~\ref{sec:Discussion}.

Even though not necessary for a proper understanding
of the present article, the reader may wish to have a look
at Appendix~A of a recent review~\citep{KV2011-review},
where the basic idea of $q$--theory is outlined.

\setcounter{equation}{0}
\renewcommand{\theequation}{\arabic{section}.\arabic{equation}}
\section{QCD-scale modified-gravity model}
\label{sec:QCD-scale-modified-gravity-model}

\subsection{Selection criteria}
\label{subsec:Selection-criteria}

There exists an infinity of $f(R)$ modified-gravity models
and an infinity thereof reproduces the main characteristics
of the observed homogeneous universe [see, e.g.,
\citet{BransDicke1961,Starobinsky1980,Bertolami1986,SotiriouFaraoni2010}].
In this paper, a subset of these theories is obtained
by imposing the following conditions:
\begin{enumerate}
  \item
Empty flat spacetime is a solution of the field equations
resulting from the action with the function $f(R)$ replaced by
the constant $f(0)$.
  \item
The action of the $f(R)$ modified-gravity model
involves only known energy scales from general relativity
and the standard model of elementary particle physics
[e.g., the energy scales
$E_\text{QCD}=\text{O}(10^2\;\text{MeV})$ and
$E_\text{Planck}\equiv \sqrt{\hbar\, c^5/(8\pi\, G_{N})}
\approx 2.44\times 10^{18}\:\text{GeV}$,
for the particular model of the next subsection].
  \item
The asymptotic de-Sitter solution from the $f(R)$ modified-gravity model
has only integer powers of the Hubble constant $H$ in the reduced action
[e.g., the terms $(E_\text{Planck})^2 H^2$ and $(E_\text{QCD})^3 H$,
for the particular model of the next subsection].
  \item
The action of the $f(R)$ modified-gravity model
has dimensionless coupling constants roughly of order
unity [e.g., the coupling constant $\eta$,
for the particular model of the next subsection].
\end{enumerate}

The first condition is motivated by the need to solve the
main cosmological constant problem [CCP1]~\citep{Weinberg1989},
namely, why is the energy scale of the effective gravitating
vacuum energy density $\rho_{V}$ negligible compared to
the basic energy scales of the standard model of elementary particle
physics (not to mention the Planck energy scale).
A first step towards solving CCP1 has been made using
$q$--theory~\citep{KV2010-CCP1}.
The third condition is to have a stationary cosmological
solution (de-Sitter universe) which is analytic
and consistent, as discussed in Section~III of \citet{KV2008-dynamics}.
The second  and fourth conditions are for simplicity's sake,
without the need to introduce new physics at ultralow energies
[that is, energies of the order of $\text{meV}$, as indicated by the
observed accelerating universe; cf.~\citet{Riess-etal1998,
Perlmutter-etal1999,Komatsu-etal2009}].\footnote{Of all four conditions,
the third has admittedly the weakest theoretical motivation.
But additional arguments may perhaps come from future results
on the quantum-field-theoretic de-Sitter state, which may hold
some surprises in store
[see, e.g., \citet{Polyakov2010} and references therein].}

Remark that the $\Lambda\text{CDM}$ model [i.e., the
cosmological model from standard general relativity with a genuine
cosmological constant $\Lambda$ and a cold-dark-matter (CDM) component]
already does not satisfy the first condition.
Moreover, the required value of $\Lambda$ in the $\Lambda\text{CDM}$
model~\citep{Weinberg1972,Carroll-etal1992,SahniStarobinsky2000,
PeeblesRatra2003,Perivolaropoulos2010}
is completely \emph{ad hoc}
(the $\text{meV}$ scale mentioned above)
and leaves CCP1 hanging in the wind,
which is theoretically unsatisfactory, as explained in, e.g.,
the second paragraph of Section~I of \citet{Mukohyama2004}.
Just to avoid any misunderstanding, the flat $\Lambda\text{CDM}$
model~\citep{Weinberg1972,Carroll-etal1992,SahniStarobinsky2000,
PeeblesRatra2003,Perivolaropoulos2010}
is perfectly satisfactory experimentally but not theoretically.
The QCD-scale modified-gravity model proposed in the next subsection
has the potential to be incorporated in a theoretically
satisfactory framework ($q$--theory) but, first, needs to
be shown  experimentally satisfactory.

\subsection{Representative model}
\label{subsec:Representative-model}

The following simplified modified-gravity
action~\citep{KV2009-gluoncond,K2010-gluoncondcosm}
satisfies the criteria of Section~\ref{subsec:Selection-criteria}
and will be the starting point of the present article:
\beqa
S_\text{grav,\,0}&=&
\int_{\mathbb{R}^4} \,d^4x\, \sqrt{-g(x)}\;
\Big[ K_{0}\,R(x)- \eta\,|R(x)|^{1/2}\,(q_{0})^{3/4}
\nonumber\\[1mm]
&&
+\mathcal{L}_{M}\big[\psi(x)\big]\Big]\,,
\label{eq:action-Sgrav0}
\eeqa
with $\hbar=c=1$ from the use of natural units,
the gravitational coupling constant $K_{0}\equiv (16\pi G_{0})^{-1}>0$,
the dimensionless coupling constant $\eta > 0$
[standard general relativity has $\eta= 0\,$], the constant
equilibrium gluon condensate $q_{0} \equiv (E_\text{QCD})^4$,
and the generalized matter field $\psi(x)$ which includes
the fields of QCD (gluons and quarks)
and further possible fields responsible for the observed
CDM component of the present universe.\footnote{As discussed in
\citet{KV2009-gluoncond}, the
vacuum energy density of the QCD fields with dynamics
governed by $\mathcal{L}_{M}$ in \eqref{eq:action-Sgrav0}
is compensated by an appropriate value $q_0$ of the
gluon-condensate $q(x)$ field, resulting in a vanishing
effective cosmological constant.
Other (non--QCD) contributions to the gravitating vacuum energy density
are perhaps canceled by the self-adjustment of similar $q$--type fields
or by a different mechanism altogether; cf. Ftn. 2 of
\citet{KV2009-gluoncond}.}
Throughout, the conventions of \citet{Weinberg1972} are used
such as the metric signature $(-+++)$, except for the Riemann tensor,
which is taken to have a further minus sign. Specifically,
the Ricci curvature scalar of a de-Sitter universe
is given by $R=12\,H^2\geq 0$ with Hubble constant $H$.

It remains to be seen if there exist other
$f(R)$ modified-gravity models which satisfy
the criteria of Section~\ref{subsec:Selection-criteria}
and describe the observed universe equally successfully.
Perhaps a model exists involving the electroweak
scale~\citep{ArkaniHamed-etal2000,KV2009-electroweak,
K2010-electroweak,K2011-electroweak}.
Returning to the energy scale of QCD, remark that the second term
of the integrand in \eqref{eq:action-Sgrav0}
corresponds to the linear term $(E_\text{QCD})^3 H$
mentioned in the second criterium of Section~\ref{subsec:Selection-criteria}
and that this linear term may be considered to be the
leading term of a power series in $H$
(higher-order terms will be discussed shortly).
Anyway, model \eqref{eq:action-Sgrav0} based on the QCD energy scale
is a perfect example (essentially unique up to higher-order terms) of a
model satisfying the criteria of Section~\ref{subsec:Selection-criteria}.
As explained in Section~\ref{sec:introduction},
the aim of the present article is to study the use of this model
as a compact description of the observed universe.

In the standard formulation of modified-gravity models
where the Ricci scalar $R$ of the Einstein--Hilbert action density
is replaced by $R+f(R)$, the proposed modification is given by
\bsubeqs\label{eq:f-L0}
\beq\label{eq:f}
f(R) = - |R|^{1/2}/L_{0} \,,
\eeq
with length scale
\beq\label{eq:L0}
L_{0}  \equiv  \eta^{-1}\,K_{0}  \,(q_{0})^{-3/4}\,.
\eeq
\esubeqs
It is important to state, right from the start, that the theory
\eqref{eq:action-Sgrav0} is only considered to be
relevant over cosmological length scales (small curvatures $|R|$).
For smaller length scales (larger curvatures  $|R|$),
other terms than the single root term \eqref{eq:f-L0}
may become important. One example would be having
$f(R)$ of \eqref{eq:f-L0} replaced by the following extended
function:
\beq\label{eq:f-ext}
\widetilde{f}_\text{ext}(R) =
- \,\frac{|R|^{1/2}/L_{0} }{1+ \zeta\,|R|^{1/2}\,L_{0}}\;,
\eeq
with another positive coupling constant $\zeta$ of order unity.
This particular function differs from the one
given by \citet{K2010-gluoncondcosm} in that its
second derivative is positive for all finite values of $R$.

\subsection{Additional remarks}
\label{subsec:Additional-remarks}

The model of the previous subsection with a constant
value $q_0$ of the gluon condensate is a simplified
version of the model with a dynamic gluon condensate field $q(x)$
considered in \citet{K2010-gluoncondcosm}.
It was found in \citep{K2010-gluoncondcosm} that the
gravitating vacuum energy density $\rho_{V}(q)\propto (q-q_0)^2$
of the corresponding cosmological model
is negligible during the late evolution of the model universe
and that the dynamic vacuum variable $q(x)$
is effectively frozen to its constant
equilibrium value $q_{0}$ with $\rho_{V}(q_{0})=0$.
For this reason, it makes sense to restrict the consideration
to model \eqref{eq:action-Sgrav0}
with just a constant value $q_0$.

The actual value of the gluon-condensate $q_{0}$
entering the nonanalytic modified-gravity term \eqref{eq:f-L0}
arises from the flat-spacetime part of the Lagrange density,
$\mathcal{L}_{M}$, and
is taken to have a standard (positive) value
$q_{0}  \approx (300\;\text{MeV})^4$;
see \citet{ShifmanVainshteinZakharov1979,Narison1996}.
Later, it will be shown that this particular value
for $q_0$ implies $\eta \approx 2\times 10^{-4}$.
Observe that, strictly speaking, the quantity $q_{0}$
of the second term in the integrand of \eqref{eq:action-Sgrav0} need
not be equal to the equilibrium value of the gluon condensate $q(x)$,
but can, in principle, be given by any typical QCD energy density.
Still, the identification of $q_{0}$ as the
equilibrium gluon condensate is maintained throughout this article,
because, with this identification, the coupling
constant $\eta$ in \eqref{eq:action-Sgrav0} is defined unambiguously.

The relation between the gravitational constant
$G_{0} \equiv G(q_{0})\equiv (16\pi K_{0})^{-1}$
from \eqref{eq:action-Sgrav0} and Newton's constant
$G_{N}$~\citep{Cavendish1798,MohrTaylorNewell2008}
is rather subtle, but, in this article, the approximate equality
$G_{0} \sim G_{N}$ is simply taken to hold
(see the next paragraph for the argument).
For these numerical values of the constants
$\eta\,(q_{0})^{3/4}$ and $G_{0}$, the length scale entering the
modified-gravity action term \eqref{eq:f-L0} has the value
$L_{0}  \sim 10^{26}\;\text{m} \sim 3\;\text{Gpc}$.

The field equations from \eqref{eq:action-Sgrav0} are fourth order
and it is advantageous to switch to the scalar-tensor formulation,
which has field equations of second order but involves an extra scalar
field~\citep{BransDicke1961,Starobinsky1980,SotiriouFaraoni2010}.\footnote{
The modified Einstein equation [see, e.g., Eq.~(6) of
\citet{SotiriouFaraoni2010}] also has singular terms
proportional to $|R|^{-1/2}$ or higher negative powers of $|R|$.
It remains to be seen whether or not this leads to unacceptable behavior.
As mentioned in~\citet{K2010-gluoncondcosm}, appropriate versions
of the model may satisfy solar-system tests because of the
chameleon effect (to be discussed shortly).
See also the related discussion in the penultimate paragraph of
Section~\ref{sec:Numerical-results}.
}
The scalar-tensor theory equivalent to \eqref{eq:action-Sgrav0}
has been given by Eq.~(2.2) of \citet{K2010-gluoncondcosm},
with $q$ replaced by $q_{0}$ and changing the sign in front of
the $\phi\,R$ term, where $\phi\in (-\infty,1)$
is the dimensionless Brans--Dicke scalar field.
The corresponding potential $U(\phi)\propto \eta^2/(1-\phi)$
is such that it leads to the
chameleon effect~\citep{KhouryWeltman2004} and
the effect is taken to be operative for Cavendish-type
experiments on Earth~\citep{Cavendish1798,MohrTaylorNewell2008},
giving $G_{0} \sim G_{N}$
(see also Endnote [39] in \citet{K2010-gluoncondcosm}
for further discussion).

\setcounter{equation}{0}
\renewcommand{\theequation}{\arabic{section}.\arabic{equation}}
\section{One-component QCD-scale modified-gravity universe}
\label{sec:One-component-modified-gravity-universe}
\vspace*{0mm}

\subsection{Pressureless matter component}
\label{subsec:From-three-components-to-one}

In order to study cosmological aspects of the modified-gravity
model \eqref{eq:action-Sgrav0},
consider a spatially flat ($k=0$) Robertson--Walker
metric~\citep{Weinberg1972} with scale factor $a(\tau)$
for cosmic time $\tau$ (later, $t$ will denote the dimensionless
quantity) and matter described by homogeneous perfect fluids.
Recall that the Hubble parameter is given by
$H(\tau) \equiv [d a(\tau)/d\tau]/a(\tau)$.

In fact, only one matter component (labeled $n=2$)
will be considered in the rest of this article, namely,
a perfect fluid of pressureless nonrelativistic matter
[e.g., cold dark matter]
with energy density $\rho_{M2}(\tau)$ and
constant equation-of-state (EOS) parameter
$w_{M2}\equiv P_{M2}/\rho_{M2}=0$.
This is a simplification of the model considered
by \citet{K2010-gluoncondcosm},
which had also a dynamic gluon-condensate component
(labeled $n=0$) with EOS parameter $w_{0}=-1$
and an ultrarelativistic-matter component
(labeled $n=1$) with EOS parameter $w_{M1}=1/3$.
In order to allow for an easy comparison with the results
of \citet{K2010-gluoncondcosm}, the label $n=2$
will be kept for the single matter component
considered in this article.

Apart from this cold-dark-matter component,
the only other input of the cosmological model
is the modified-gravity term
from \eqref{eq:action-Sgrav0}. Using the scalar-tensor formalism,
there is then the auxiliary Brans--Dicke scalar $\phi(\tau)$
to consider, without direct kinetic term but with the nontrivial
potential $U(\phi)$ already mentioned in
Section~\ref{subsec:Additional-remarks}.

\subsection{Dimensionless variables and ODEs}
\label{subsec:Dimensionless-equations}

The following dimensionless variables
$t $, $h$, $f$, $u$, $s$, and $r$ can be introduced:%
\bsubeqs\label{eq:Dimensionless-var}
\beqa
\tau   &\equiv& t  \;K_{0}\big/(\eta\,(q_{0})^{3/4})\,,\\[2mm]
H(\tau)&\equiv& h(t)\;\eta\,(q_{0})^{3/4}\big/K_{0}\,,\\[2mm]
U(\tau)   &\equiv& u(t)\;\eta^2\,(q_{0})^{3/2}\big/K_{0}^2\,,\\[2mm]
\phi(\tau)&\equiv& s (t)\,,\\[2mm]
\rho_{M2}(\tau) &\equiv& r_{M2}(t)\;\eta^2\,(q_{0})^{3/2}\big/K_{0}\,,
\eeqa
\esubeqs
where, different from \citet{K2010-gluoncondcosm}, the rescaling
is done with the combination
$\eta\,(q_{0})^{3/4}/K_{0} = \eta\,(q_{0})^{3/4}\,(16\pi G_{0})$
appearing in the modified-gravity term \eqref{eq:f}.
All dimensionless quantities are denoted by lower-case Latin letters.

 From the action \eqref{eq:action-Sgrav0} in the scalar-tensor
formulation~\citep{K2010-gluoncondcosm}, there is the following
closed system of $4$ first-order
ordinary differential equations (ODEs) for the $4$
dimensionless variables  $h(t)$, $s (t)$, $v(t)$, and $r_{M2}(t)$:%
\bsubeqs\label{eq:4ODEsFRWdim}
\beqa
\hspace*{-5mm}
\dot{h} &=& -2\,h^2 - \frac{1}{6}\,\frac{\partial u}{\partial s}\,,
\label{eq:4ODEsFRWdim-h}\\[2mm]
\hspace*{-5mm}
\dot{s} &=& v\,,
\label{eq:4ODEsFRWdim-s}\\[2mm]
\hspace*{-5mm}
\dot{v} &=&
\frac{1}{6}\,r_\text{M2}-3\,h\,v-  \frac{2}{3}\,u
+ \frac{1}{3}\,s \,\frac{\partial u}{\partial s }\,,
\label{eq:4ODEsFRWdim-v}\\[2mm]
\hspace*{-5mm}
\dot{r}_{M2}
&=&
-3\,h \,r_{M2}\,,
\label{eq:4ODEsFRWdim-rM}
\eeqa \esubeqs
where the overdot stands for differentiation with respect to
the dimensionless cosmic time $t$   and
and the dimensionless Brans--Dicke potential $u$  is given by
\beq
u(t)       =     -\frac{1}{4}\;\frac{1}{1-s (t)}\,.
\label{eq:dimensionless-potential-u}
\eeq
With the solution of the above ODEs for appropriate boundary conditions,
it is possible to verify \emph{a posteriori} the
Friedmann-type equation,
\beq
 h^2\,s +h\,v = \big(r_\text{M2} - u\big)\big/ 6  \,,
\label{eq:Friedmann-type-eq}
\eeq
which, in general, is guaranteed to hold by the contracted Bianchi
identities and energy-momentum conservation~\citep{Weinberg1972}.
Hence, if the solution of ODEs~\eqref{eq:4ODEsFRWdim}
satisfies \eqref{eq:Friedmann-type-eq} at one particular time,
then \eqref{eq:Friedmann-type-eq} is satisfied at all times.

The boundary conditions for ODEs~\eqref{eq:4ODEsFRWdim}
are obtained by setting $t=t_\text{start}$ in the following
functions:
\bsubeqs\label{eq:4approxsol}
\beqa
\hspace*{-5mm}
h^\text{approx}(t) &=& \frac{2}{3}\,\frac{1}{t}\,
\left(1+\frac{3\,\sqrt{3}}{16}   \;t - \frac{405}{512}\; t^2 \right)\,,
\label{eq:4approxsol-h}\\[2mm]
\hspace*{-5mm}
s^\text{approx}(t) &=& \left(1 - \frac{\sqrt{3}}{4}\; t + \frac{9}{16}\; t^2\right)\,,
\label{eq:4approxsol-s}\\[2mm]
\hspace*{-5mm}
v^\text{approx}(t) &=& \dot{s}^\text{approx}(t)\,,
\label{eq:4approxsol-v}\\[2mm]
\hspace*{-5mm}
r_{M2}^\text{approx}(t)&=& \frac{8}{3}\,\frac{1}{t^2}\,
\left(1- \frac{3\sqrt{3}}{8}\; t\right)\,.
\label{eq:4approxsol-rM}
\eeqa \esubeqs
These functions provide, in fact,
an approximate solution of the ODEs~\eqref{eq:4ODEsFRWdim} and
generalized Friedmann equation~\eqref{eq:Friedmann-type-eq}
in the limit $t \to 0$.

Purely mathematically, remark that the ODEs \eqref{eq:4ODEsFRWdim} have no
free parameters and that the boundary conditions are fixed completely by
\eqref{eq:4approxsol} for sufficiently small $t_\text{start}$.
Physically, however, $t_\text{start}$ must be small enough
but still larger than the time $t_\text{cross}$ corresponding
to the QCD crossover at a temperature
$T_\text{cross} \sim E_\text{QCD} \sim 300\;\text{MeV}$
[see \citet{K2010-gluoncondcosm} for further discussion].
Specifically, one must take a $t_\text{start}$ value obeying
$1 \gg t_\text{start} \gg t_\text{cross}\sim \eta\,E_\text{QCD}/E_\text{Planck}
\sim 10^{-23}$ for $\eta \sim 10^{-4}$.

\subsection{Observables quantities}
\label{subsec:Observables-quantities}

In order to test the cosmology resulting from QCD-scale modified-gravity
model \eqref{eq:action-Sgrav0}, four observable quantities will be
considered~\citep{K2010-gluoncondcosm,Zhang-etal2007}: $t_{p}\,h(t_{p})$,
$\overline{w}_{X}(t_{p})$, $z_\text{inflect}(t_{i},\, t_{p})$,
and $E_{G}^\text{\,theo}(z)$,
where $z$ stands for the redshift with respect to the present
epoch $t=t_{p}$ to be defined below. As a preliminary, note
that the generalized Friedmann equation \eqref{eq:Friedmann-type-eq} gives
\bsubeqs\label{eq:omegabar} \beqa
\overline{\omega}_{X}+\overline{\omega}_{M2}&=&1\,,
\label{eq:omegabar-XplusM}\\[1.75mm]
\overline{\omega}_{X}   &\equiv& r_{X}/(6\,h^2) \equiv
\overline{\omega}_{V}+\overline{\omega}_\text{grav}\,,
\label{eq:omegabar-X}\\[1.75mm]
\overline{\omega}_{V}&\equiv& r_{V}/(6\,h^2) =0\,,
\label{eq:omegabar-V}\\[1.75mm]
\overline{\omega}_\text{grav}&\equiv& 1-s-\dot{s}/h-u/(6\,h^2)\,,
\label{eq:omegabar-grav}\\[1.75mm]
\overline{\omega}_{M2} &\equiv& r_{M2}/(6\,h^2)\,.
\label{eq:omegabar-M}
\eeqa
\esubeqs
Without genuine vacuum energy density [$r_V(q)
\equiv K_{0} \,\eta^{-2}\,(q_{0})^{-3/2}\,\rho_{V}(q) =0$ for $q=q_{0}$],
the only new ingredient in \eqref{eq:omegabar} is
$\overline{\omega}_\text{grav}$, as it
vanishes for the standard theory with $u=0$ and $s=1$.
Remark that $r_{X}$ of \eqref{eq:omegabar-X} does not correspond
to a real physical energy density but is a mathematical quantity
inferred from the variables $\overline{\omega}_{V}$
and $\overline{\omega}_\text{grav}$ defined by
\eqref{eq:omegabar-V} and \eqref{eq:omegabar-grav}.

The energy density parameters of \eqref{eq:omegabar} are defined
in terms of the gravitational constant $G_{0} \equiv G(q_{0})$,
which may, in principle, be different from
Newton's constant $G_{N}$ as measured by laboratory experiments on
Earth~\citep{Cavendish1798,MohrTaylorNewell2008}.
For this reason, the parameters have been denoted by
$\omega$ and not by the standard symbol $\Omega$.
Still, the precise relation between $G_{0}$  and $G_{N}$
will be important only once in Section~\ref{sec:Numerical-results}, namely,
when the absolute age of the universe is calculated and, then,
the equality $G_{0}=G_{N}$ is taken to hold [assuming the chameleon
effect~\citep{KhouryWeltman2004} to be relevant for earth-based experiments,
as discussed in Section~\ref{subsec:Additional-remarks}].
Now, turn to the four observables mentioned above.

The first observable is simply the expansion rate of the universe
in units of the inverse of its age,
\beq\label{eq:tphp}
t_{p}\,h(t_{p}) \equiv t\, \dot{a}(t)/a(t)\,\big|_{t=t_{p}}\,,
\eeq
where $a(t)$ is the scale factor of the Robertson--Walker
metric. This observable quantity has been evaluated for the ``present epoch,''
which is taken to be
defined by the moment $t=t_{p}$ when $\overline{\omega}_{M2}(t_{p})=1/4$
or $\rho_{X}(\tau_{p})/\rho_{M2}(\tau_{p})=3$
from \eqref{eq:omegabar-XplusM}.
The fiducial value $\overline{\omega}_{M2}(t_{p})=1/4$ will be used
in the following and is more or less consistent with the
available data [see, e.g., \citet{Komatsu-etal2009}].

The second observable corresponds to the effective EOS parameter
of the unknown component $X$, whose model value can be extracted
from \eqref{eq:4ODEsFRWdim} and \eqref{eq:Friedmann-type-eq}:
\beqa\label{eq:modgrav-wXbar-tp}
\overline{w}_{X}(t_{p})
&\equiv&
-\frac{2}{3}\,\left(\frac{\ddot{a}\,a}{(\dot{a})^2}+\frac{1}{2}\right)\;
 \frac{1}{1-\overline{\omega}_{M2}}\;\Bigg|_{t=t_{p}}
\nonumber\\[2mm]
&=&
 -\;\frac{  u +4\,h\,\dot{s} +2\,\ddot{s} \phantom{\;(1-s)}}
         {\,u +6\,h\,\dot{s} -r_{M2}\,(1-s)}\;\Bigg|_{t=t_{p}}\;.
\eeqa
Again, this observable quantity has been evaluated at the present
epoch $t=t_{p}$. For later times,
the right-hand side of \eqref{eq:modgrav-wXbar-tp} shows that
$\overline{w}_{X}$ of the modified-gravity model \eqref{eq:action-Sgrav0}
approaches the value $-1$ in the limit of vanishing
matter content $r_{M2}$ and constant Brans--Dicke scalar $s$
as $t\to\infty$.

The third type of observable follows from
the transition of deceleration to acceleration.
In mathematical terms, this time corresponds
to the nonstationary inflection point of the function $a(t)$, that is,
the value $t_{i}$ at which the second derivative of $a(t)$ vanishes
but not the first derivative. Specifically, the inflection point
$t=t_{i}$  corresponds to the following redshift for an observer
at $t=t_{p}\geq t_{i}\,$:
\beq
z_\text{inflect}(t_{i},\, t_{p}) \equiv a(t_{p})/a(t_{i})-1 \,.
\label{eq:z_inflect}
\eeq

In order to prepare for the fourth observable of the empirical
model \eqref{eq:action-Sgrav0}, turn to the linear growth of
sub-horizon matter-density perturbations
in the Newtonian gauge~\citep{Song-etal2007}.
For small enough matter-density-perturbation amplitudes
$\Delta_{M2} \equiv \delta r_{M2}/r_{M2}$
and large enough wavelengths (but still within the horizon),
the following dimensionless linear ODE needs to be solved:
\beq\label{eq:ODE-DeltaM2}
\ddot{\Delta}_{M2}(t) +2\, h(t)\, \dot{\Delta}_{M2}(t)
-\frac{1}{3}\,\frac{1}{s(t)}\,r_{M2}(t)\,\Delta_{M2}(t)
=0 \,,
\eeq
which is equivalent to, for example,
Eq.~(13) of \citet{Tsujikawa-etal2009}
or Eq.~(15) of \citet{Lombriser-etal2010},
with an extra  factor $(4/3)\,(1/\phi)$ multiplying the original
non-derivative term.
See also \citet{delaCruzDombriz-etal2008}, where the approximate
ODE \eqref{eq:ODE-DeltaM2}
has been tested for an $|R|^{1/2}$ modified-gravity model.

 From the combined solutions for $a(t)$ and $\Delta_{M2}(t)$, the
following linear growth parameter is obtained by differentiation:
\beq\label{eq:beta}
\beta \equiv   \frac{d \ln \Delta_{M2}}{d \ln a}
       =       \frac{1}{h}\,\frac{\dot{\Delta}_{M2}}{\Delta_{M2}}\,.
\eeq
For the early phase with  $\phi=s\sim 1$, $h\sim 2/3\,t^{-1}$,
and $r_{M2}\sim 6\,h^2$,
the linear growth parameter calculated from \eqref{eq:ODE-DeltaM2} is
\beq\label{eq:beta-early}
\beta_\text{early} = \big(\sqrt{33}-1\big)\big/4 \approx 1.186\,.  
\eeq
The corresponding matter-density perturbation is given by
\beq\label{eq:DeltaM2-early}
\Delta_{M2}^\text{approx}(t)\propto t^{\,2\beta_\text{early}/3}\,,
\eeq
which provides an approximate solution of the ODE~\eqref{eq:ODE-DeltaM2}
in the limit $t \to 0$.

Now, consider the estimator $E_{G}$ introduced by
\citet{Zhang-etal2007} with the purpose of searching for
deviations from Einstein gravity.
According to the original paper of \citet{Zhang-etal2007}
and also the paper of \citet{Lombriser-etal2010},
the theoretical expression for
$E_{G}(z)$ in the modified-gravity model \eqref{eq:action-Sgrav0}
is
\beq\label{eq:E_G-theo}
E_{G}^\text{\,theo}(z) =\frac{\overline{\omega}_{M2}(t_{p})}{(1+f_R(z))\,\beta(z)}
                  =\frac{\overline{\omega}_{M2}(t_{p})}{\phi(z)\,\beta(z)}\,,
\eeq
with $f_R\equiv d f/d R$.
The last expression of \eqref{eq:E_G-theo} is given in terms of
the scalar-tensor formalism, where the relevant
field equations~\citep{K2010-gluoncondcosm} have been used. From the
numerical solution of the ODEs to be presented
in Section~\ref{sec:Numerical-results},
the values of $\phi(z)$ and $\beta(z)$ are readily obtained
and, thereby, the numerical value of $E_{G}^\text{\,theo}(z)$.
In the early phase with $\phi\sim 1$ and $\beta$ given by \eqref{eq:beta-early},
one has
\beq\label{eq:E_G-theo_early}
E_{G}^\text{\,theo}\big|_{z \gg 1} \sim
\overline{\omega}_{M2}(t_{p})/\beta_\text{early} \approx 0.2108\,,
\eeq
for the fiducial value $\overline{\omega}_{M2}(t_{p})=0.25$
mentioned earlier.

\setcounter{equation}{0}
\renewcommand{\theequation}{\arabic{section}.\arabic{equation}}
\section{Benchmark results from the $\boldsymbol{\Lambda\text{CDM}}$ model}
\label{sec:Benchmark-LambdaCDM-results}

Four observables of the QCD-scale modified-gravity
model \eqref{eq:action-Sgrav0} have been discussed
in  Section~\ref{subsec:Observables-quantities} and will
be evaluated numerically in Section~\ref{sec:Numerical-results}.
In this section, the corresponding values are given
for the spatially flat $\Lambda\text{CDM}$--model
universe~\citep{Weinberg1972,Carroll-etal1992,SahniStarobinsky2000,
PeeblesRatra2003,Perivolaropoulos2010}.

Recall that the relevant spatially flat $\Lambda\text{CDM}$--model
is completely defined by the value of the cosmological
constant $\Lambda$ and the condition that $\Omega_{M2}$ equals
$1/4$ at present. Here, the standard matter energy-density
parameter in terms of Newton's constant
$G_{N}$ is given by \citet{Weinberg1972}
\bsubeqs
\beqa
\Omega_{M2}(\tau_{p})            &\equiv&
\rho_{M2}(\tau_{p})/\rho_{\text{crit}}(\tau_{p})\,,\\[2mm]
\rho_{\text{crit}}(\tau_{p}) &\equiv& 3\,H^2(\tau_{p})/(8\pi\,G_{N})\,,
\eeqa
\esubeqs
where the present epoch occurs at cosmic time $\tau=\tau_{p}$.
As mentioned in Section~\ref{subsec:Selection-criteria},
the $\Lambda\text{CDM}$ model may be theoretically unsatisfactory
but has been found to give an excellent description
of the observed accelerating
universe~\citep{Riess-etal1998,Perlmutter-etal1999,Komatsu-etal2009}.
For this reason, it can provide benchmark results to compare
other models with.

Analytic results for the first three observables
of Section~\ref{subsec:Observables-quantities} are
as follows~\citep{K2010-gluoncondcosm}:
\beqa\label{eq:3obs-LCDM}
&&
\Big(
\tau_{p}\,H(\tau_{p}),\,
\overline{w}_{X}(\tau_{p}),\,
z_\text{inflect}(\tau_{i} ,\, \tau_{p})
\Big)
\,\Big|_{\Lambda\text{CDM}}
\nonumber\\[1mm]
&&=
\Big(
(4\, \text{arcsinh}\,\sqrt{3})/(3\,\sqrt{3}),\,
-1,\,
(6^{1/3}-1)
\Big)
\nonumber\\[1mm]
&&\approx
\Big( 1.01,\,   -1,\,  0.817 \Big)\,. 
\eeqa
The fourth observable is simply~\citep{Zhang-etal2007}
\beq\label{eq:E_G-theo-LCDM}
E_{G}^\text{\,theo}(z)\,\big|_{\Lambda\text{CDM}}
=\Omega_{M2}(\tau_{p})/\beta_{\Lambda\text{CDM}}(z)  \,,
\eeq
where $\beta_{\Lambda\text{CDM}}(z)$ can be calculated analytically
from the exact solution of the linear ODE corresponding to \eqref{eq:ODE-DeltaM2};
see, in particular, the $C_2$ term in Eq.~(6.67) of
\citet{Mukhanov2005}.
The estimator \eqref{eq:E_G-theo-LCDM} approaches the
constant value $\Omega_{M2}(\tau_{p})$ for increasing redshift $z$,
as $\beta_{\Lambda\text{CDM}}(z)$ goes to $1$ for $z\to\infty$.

\setcounter{equation}{0}
\renewcommand{\theequation}{\arabic{section}.\arabic{equation}}
\section{Numerical results for the modified-gravity universe}
\label{sec:Numerical-results}

\begin{figure}[p]
\begin{center} 
\epsfig{file=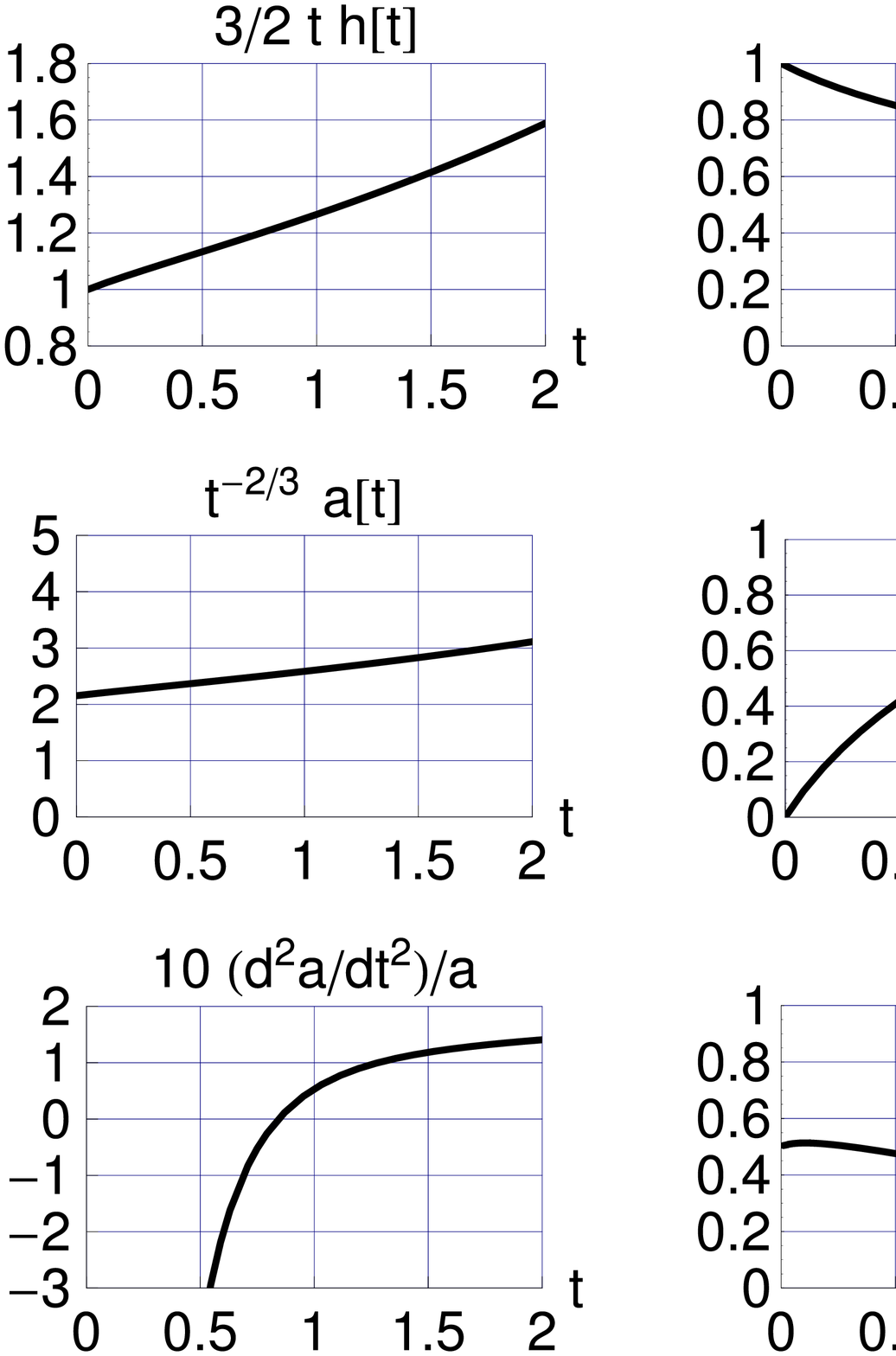,
        width=0.9\textwidth}
\end{center}
\vspace*{1mm}
\caption{Numerical solution of the modified-gravity cosmological
ODEs~\eqref{eq:4ODEsFRWdim}, with Brans--Dicke scalar
potential \eqref{eq:dimensionless-potential-u} and
pressureless matter having a dimensionless energy density $r_{M2}$
[see \eqref{eq:Dimensionless-var} for the definitions of
the dimensionless variables used]. The boundary conditions follow
from the approximate solution \eqref{eq:4approxsol} evaluated
at $t _\text{start}=10^{-5}$.
The initial value $a(t _\text{start})$ for the scale factor $a(t)$
is taken as $10^{-3}$ but has no direct physical relevance. The linear
matter-density-perturbation ODE \eqref{eq:ODE-DeltaM2} is solved
simultaneously with boundary conditions from \eqref{eq:DeltaM2-early}
evaluated at $t_\text{start}$.
The figure panels are organized as follows: the panels of the
first column from the left concern the scale factor $a(t)$
and the Hubble parameter $h\equiv (d a/d t)/a$, those of the
second column the Brans--Dicke scalar $s(t)$
[the QCD-modified-gravity model \eqref{eq:action-Sgrav0}
being studied in the scalar-tensor formalism],
those of the third column the matter energy density $r_{M2}$
[the bottom panel of this column showing the linear growth parameter
$\beta$ of subhorizon matter-density perturbations],
and those of the fourth column derived quantities
[the bottom panel showing the gravity estimator $E_{G}^\text{\,theo}\,$
defined by \eqref{eq:E_G-theo}].
The three panels of the fourth column can be combined
to give the behavior of $h(z)/h(0)$ and $E_G(z)$
shown in Table~\ref{tab-H-EG-results}.
The several energy-density parameters $\overline{\omega}$
and the effective ``dark-energy''
equation-of-state parameter $\overline{w}_{X}$ are defined in
\eqref{eq:omegabar} and \eqref{eq:modgrav-wXbar-tp}, respectively.
\vspace*{6mm}}
\label{fig:1}
\end{figure}

Figure~\ref{fig:1} displays the numerical solution
of the first-order nonlinear ODEs \eqref{eq:4ODEsFRWdim}
with boundary conditions  \eqref{eq:4approxsol}
at $t=t _\text{start}=10^{-5}$.
The simultaneous numerical solution of the second-order linear ODE
\eqref{eq:ODE-DeltaM2} has been obtained for
boundary conditions from the approximate solution
\eqref{eq:DeltaM2-early} at $t=t _\text{start}$.
Specifically, \eqref{eq:DeltaM2-early} provides the initial derivative
$\dot{\Delta}_{M2}(t_\text{start})$ for a given initial value of
$\Delta_{M2}(t_\text{start})$.
Numerical solutions of \eqref{eq:ODE-DeltaM2} with other
values for the initial derivative $\dot{\Delta}_{M2}(t_\text{start})$
have been seen to rapidly approach the approximate solution
\eqref{eq:DeltaM2-early}, provided the dimensionless cosmic time $t$
remains small enough.

\begin{figure}
\begin{center}   
\hspace*{-5mm}
\epsfig{file=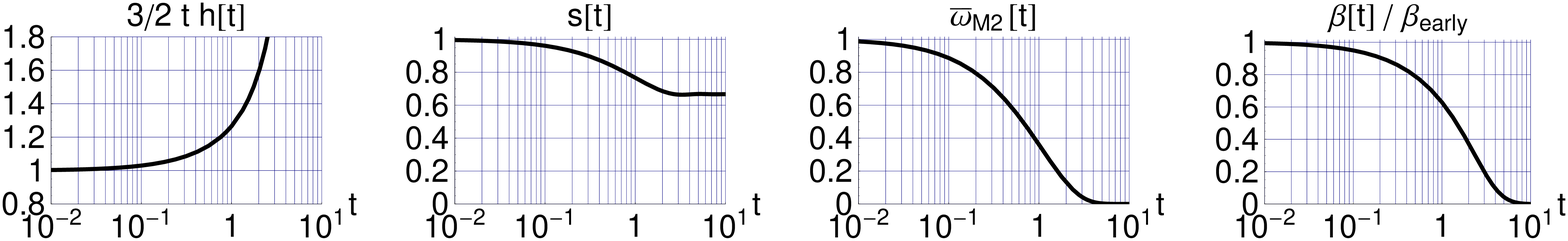,
        width=1.05\textwidth}
\end{center}
\vspace*{1mm}
\caption{Semi-log plot of numerical quantities
from Fig.~\ref{fig:1}, with
$\beta_\text{early}$ defined by \eqref{eq:beta-early}.}%
\label{fig:2}
\end{figure}

The linear $t$ scale of Fig.~\ref{fig:1} is
convenient for the late evolution of the model universe,
because, as will be seen shortly, $t$ corresponds to the
cosmic time $\tau$ measured in units of approximately
$10^{10}\;\text{yr} = 10\;\text{Gyr}$.
A logarithmic scale is, however, more appropriate for the early phase
and Fig.~\ref{fig:2} shows that the numerical results
from the QCD-scale modified-gravity model \eqref{eq:action-Sgrav0}
reproduce the Friedmann--Robertson--Walker--type expansion with
Hubble parameter $h=2/3\,t^{-1}$ and linear growth parameter $\beta$
given by  \eqref{eq:beta-early}.
Table~\ref{tab-tmin-results}
displays, moreover, a stable behavior of the $t=t_{p}$ observables
within the numerical accuracy
(at the one-per-mill level or better, for the quantities shown).

The results for the first three observable quantities
of Section~\ref{subsec:Observables-quantities} are as follows:
\beqa\label{eq:3obs-QCDmodgrav}
&&
\Big(
\tau_{p}\,H(\tau_{p}),\,
\overline{w}_{X}(\tau_{p}),\,
z_\text{inflect}(\tau_{i} ,\, \tau_{p})
\Big)
\nonumber\\[1mm]
&&
\approx
\Big( 0.917,\,  -0.662,\,  0.523 \Big)\,.
\eeqa
These results may be compared to the $\Lambda\text{CDM}$--model
values \eqref{eq:3obs-LCDM}.

The obtained values \eqref{eq:3obs-QCDmodgrav} are
consistent with those of Table~I in \citet{K2010-gluoncondcosm}
for the modified-gravity model with a dynamic $q$ field,
but without the need to consider the limit of the mathematical
parameter $Z \equiv  (q_{0})^{1/2}\;K_{0} ^{-1}
             \equiv  2\,(E_\text{QCD}/E_\text{Planck})^2$
to a numerical value of order $10^{-38}$,
as $Z$ has been scaled away completely in the
present simplified model.
The only place where this hierarchy parameter $Z$ enters is for
the dimensional age of the present universe and its expansion rate.

In fact, with $G_{0} =G_{N}$ and the elementary-particle-physics
result~\citep{ShifmanVainshteinZakharov1979,Narison1996}
for the flat-spacetime gluon condensate
$q_{0}$ $\equiv$ $(E_\text{QCD})^4$ $=$ $(300\;\text{MeV})^4$,
the values
$(t_{p},\,h_{p})$ $=$ $(1.374,$ $0.6673)$  
from the numerical solution (Table~\ref{tab-tmin-results})
give the following results for the present age
and expansion rate of the model universe:
\bsubeqs\label{eq:results-taup-Hp-eta-estimate}
\beqa
\tau_{p}
&=& t_{p}\,\eta^{-1}\,K_{0} \,(q_{0})^{-3/4}
\nonumber\\[1mm]
&=&   t_{p}\,\eta^{-1}\,(16\pi G_{N})^{-1}\,(E_\text{QCD})^{-3}
\nonumber\\[1mm]
&\approx& 13.2\;\text{Gyr} \,, 
\label{eq:results-tp}\\[2mm]
H_{p}
&=& h_{p}\,\eta\,K_{0}^{-1}\,(q_{0})^{3/4}
\nonumber\\[1mm]
&=&   h_{p}\,\eta\,(16\pi G_{N})\,(E_\text{QCD})^{3}
\nonumber\\[1mm]
&\approx& 68.1 \;\text{km}\;\text{s}^{-1}\;\text{Mpc}^{-1}\,,  
\label{eq:results-Hp}\\[2mm]
\eta &\approx& 2.40 \times 10^{-4}\,,
\label{eq:eta-estimate}
\eeqa
\esubeqs
where the numerical values given in
\eqref{eq:results-tp} and \eqref{eq:results-Hp}
use the  $\eta$ value given in \eqref{eq:eta-estimate}.
The modified-gravity model \eqref{eq:action-Sgrav0} can thus
give reasonable values for \emph{both} the present age and expansion
rate, which traces back to the fact that the numerical value for
the product $t_{p}\,h_{p}$ has been found to be close to 1.
Only the actual value of $\eta$ depends on the precise determination
of $q_{0}$ and $G_{0}/G_{N}$.
All other results of this paper do not depend on  $\eta$ , $q_{0}$,
or $G_{0}$ as these quantities can be scaled out according
to \eqref{eq:Dimensionless-var}. The rest of this section focusses
on two such observable quantities, $H(z)/H(0)$ and $E_{G}(z)$,
both considered as a function of the redshift $z$.

The behavior of the Hubble parameter $H(z)$ relative to the calculated
present value \eqref{eq:results-Hp} is given by the second column of
Table~\ref{tab-H-EG-results}. The corresponding modified-gravity results
for $H(z)$ agree with the observations reported in Fig.~9 and Table~2
of \citet{Stern-etal2010}, even though the scatter of the data
is still substantial. The third column of
Table~\ref{tab-H-EG-results} gives, for comparison,
the Hubble-parameter ratios from the $\Lambda\text{CDM}$ model,
which also agree with the current $H(z)$ observations~\citep{Stern-etal2010}
if the measured value of $H(0)$ is used as input.

The numerical solution of Fig.~\ref{fig:1} at $t=1.0011$ gives
the redshift $z \approx 0.32$,
the scalar-field value $\phi=s \approx 0.7666$,
and the linear growth parameter $\beta \approx 0.7459$.
These numbers combined with the fiducial value $\overline{\omega}_{M2p}=0.25$
result in the following numerical value
for the gravity estimator \eqref{eq:E_G-theo}:
\beq\label{eq:E_G-estimate}
E_{G}^\text{\,theo}\,\big|_{z = 0.32}\approx 0.437\,,  
\eeq
which agrees within one sigma with the experimental
result $E_{G}^\text{\,exp}=0.392 \pm 0.065$
from \citet{Reyes-etal2010} for a sample of galaxies
with an average redshift $\langle z \rangle =0.32$.

The same quantity $E_{G}^\text{\,theo}$ at $z\approx 0.3$ has also been
calculated by \citet{Lombriser-etal2010}
for an $f(R)$ modified-gravity model  with $1+f_{R}=\phi\sim 1$ and
$B_{0} $ parameter~\citep{Song-etal2007} of order $0.2$,
giving a value around $0.35$. In order to compare to this result,
the value of the $B_{p}$ parameter
(in the notation with the present epoch at $\tau=\tau_{p}$)
has been calculated for the QCD-scale modified-gravity
model \eqref{eq:action-Sgrav0}:  \vspace*{-0mm}
\beqa\label{eq:B0-def-estimate}
B_{p}
&\equiv&
\frac{R\, f_{RR}}{1+f_R}\,
\bigg(\frac{1}{R}\;\frac{d R}{d \ln a}\bigg)\,\bigg(H\,\frac{d \ln a}{d H}\bigg)
\;\bigg|_{\tau=\tau_{p}}
\nonumber\\[1mm]
&=&
\frac{1}{2}\;\frac{1}{2\, L_{0}  \sqrt{|R|}-1}\,
\bigg(\frac{\dot{R}}{R}\;\frac{H}{\dot{H}}\bigg)
\;\bigg|_{\tau=\tau_{p}}
\nonumber\\[1mm]
&\approx& 0.246\,,  
\eeqa
where the Ricci scalar is given by $R=6\,(d H/d\tau+2\,H^2)$
and the function $f(R)$ is defined by \eqref{eq:f-L0}.
With this $B_{p}$ parameter and simply omitting the factor $1/(1+f_R)$
or $1/\phi$ in \eqref{eq:E_G-theo},
our result for $E_G$ at z = 0.32 would be approximately $0.34$,
which would agree with the result of
Fig.~4 in \citet{Lombriser-etal2010}.

Nevertheless, the factor $1/(1+f_R)$  is unarguably present in
the $E_G$ expression \eqref{eq:E_G-theo}
and the correct estimate from the QCD-modified-gravity model
is \eqref{eq:E_G-estimate}, which is
larger than that of \citet{Lombriser-etal2010} but still
consistent with the direct measurement~\citep{Reyes-etal2010}.
Observe that the value $f_R(t_{p}) \approx 0.7259 -1 \approx -0.2741$
is consistent with
the cosmic-microwave-background data according to Table~III
of \citet{Lombriser-etal2010},
but apparently inconsistent with the cluster-abundance data
according to Table~IV of the same reference.
However, as mentioned before, the QCD-modified-gravity model
is assumed to hold only for the very largest length scales
and not galaxy-cluster length scales
(or, \emph{a forteriori}, solar-system length scales).
Interestingly, the cluster-abundance data can be used to
constrain the extensions of the simple $|R|^{1/2}$ term,
one possibility having been mentioned in \eqref{eq:f-ext}.

Finally, additional  values for $E_{G}^\text{\,theo}$ at selected
redshifts are given in the fourth column of
Table~\ref{tab-H-EG-results}. This table also compares
the QCD-modified-gravity results for $E_{G}^\text{\,theo}$
with those of the $\Lambda\text{CDM}$--model universe.
[The $\Lambda\text{CDM}$ value for $E_{G}^\text{\,theo}$
at $z = 0.32$ is $0.396$, whereas the modified-gravity value
has already been given in \eqref{eq:E_G-estimate}
and the experimental value just below that equation.]
Putting the theoretical predictions for $E_{G}(z)$
in Table~\ref{tab-H-EG-results} next to
the simulated data in Fig.~1 of \citet{Zhang-etal2007}
suggests that it may be difficult for future surveys
(e.g., the Square Kilometer Array on the ground or the
Joint Dark Energy Mission and the Euclid Satellite in space)
to distinguish between the two theoretical models
but perhaps not impossible.

\setcounter{equation}{0}
\renewcommand{\theequation}{\arabic{section}.\arabic{equation}}
\section{Discussion}
\label{sec:Discussion}

In this article, a simple empirical model has been proposed
with a QCD-scale modified-gravity term \eqref{eq:action-Sgrav0} and
a single pressureless matter component (cold dark matter, CDM).
This particular $f(R)$ modified-gravity model
has been selected on physical grounds
(Section~\ref{subsec:Selection-criteria}), but is, in the end,
solely used as an efficient
way to describe the main aspects of the late evolution of the universe,
having only two fundamental energy scales, $E_\text{QCD}$ and
$E_\text{Planck}$, and a single dimensionless coupling constant, $\eta$.

With the elementary-particle-physics value for the equilibrium
gluon condensate $q_{0} \equiv (E_\text{QCD})^4 = (300\;\text{MeV})^4$,
the measured age of the universe fixes the dimensionless
coupling constant $\eta$ of model \eqref{eq:action-Sgrav0} to the
value of approximately $2.4 \times 10^{-4}$.
As emphasized in \citet{K2010-gluoncondcosm},
the effective coupling constant $\eta$ may ultimately be calculated
from QCD and general relativity; see also
\citet{Schuetzhold2002,Bjorken2004,UrbanZhitnitsky2010a,
UrbanZhitnitsky2010b,Holdom2011}
for further discussion of the possible relation of QCD and dark energy.
The connection of the QCD-scale modified-gravity model with the
$q$--theory approach to solving the main cosmological constant problem
has already been mentioned in Section~\ref{subsec:Selection-criteria}.

The $q_{0}$ and $\eta$ values quoted in the previous paragraph
result in a model postdiction for the Hubble constant \eqref{eq:results-Hp},
which is within $10\,\%$ of the observed value. Theoretically,
the last mathematical expression for the Hubble constant in the middle
of \eqref{eq:results-Hp} is quite remarkable, as it involves
only Newton's constant $G_{N}$ and the cube of the energy
scale $E_\text{QCD}$.
Independent of the $q_{0}$ and $\eta$ values, there are further
model predictions for the dimensionless quantities \eqref{eq:3obs-QCDmodgrav}
and \eqref{eq:E_G-estimate}, which are again in the same ball park as
the observed values. The modified-gravity model prediction of $-0.7$
for the quantity $\overline{w}_{X}(\tau_{p})$
as defined by \eqref{eq:modgrav-wXbar-tp}
[compared to the $\Lambda\text{CDM}$ value of $-1$]
may perhaps provide a crucial test, as long as
\emph{independent} measurements of the present values
of $H$, $d H/d\tau$, and $\rho_\text{CDM}\equiv \rho_\text{M2}$
can be obtained.

Moreover, model predictions for the redshift dependence of the gravity
estimator $E_{G}$ have been given in Table~\ref{tab-H-EG-results}.
At the very largest length scales (comoving wavelengths of the order
of hundred Mpc or more but still less than the horizon scale)
and relatively low redshifts $z \sim 0.75$,
the QCD-modified-gravity results for $E_{G}$ differ by some
$+10\%$ from the  $\Lambda$CDM--model values.
The QCD-modified-gravity results for $E_{G}$ differ by some
$-10\%$ from the $\Lambda$CDM--model values for $z \gg 3$, but it is
not yet clear how these redshifts can be probed observationally.

As it stands, the QCD-scale modified-gravity model
\eqref{eq:action-Sgrav0} gives a remarkable description of
the main aspects of the late evolution of the universe.
With many forthcoming ground-based experiments and space missions,
future measurements of $\overline{w}_{X}(\tau_{p})$ and $E_{G}(z)$
may suggest alternative $f(R)$ functions,
such as the one given by \eqref{eq:f-ext}.
Of course, if experiment indeed finds evidence for nonzero $f(R)$,
it is up to theory to provide the proper understanding.

\section*{Acknowledgments}

The author thanks S. Appleby and J. Weller for pointing out
a problem with an earlier version of Eq.~\eqref{eq:f-ext}
and both ASR referees for constructive comments.
He also gratefully acknowledges the hospitality of the IPMU
during the month of May 2010.
This work was supported in part by the
World Premier International Research Center Initiative
(WPI Initiative), MEXT, Japan.


\vfill

\begin{table}[p]
\caption{Function values for the ``present epoch''
[defined by $\overline{\omega}_{M2}(t_{p})=0.25$] from the numerical
solution of the modified-gravity cosmological ODEs~\eqref{eq:4ODEsFRWdim},
with Brans--Dicke scalar potential \eqref{eq:dimensionless-potential-u}
and boundary conditions \eqref{eq:4approxsol} taken at
different values of $t_\text{start}$.
The values for the linear growth parameter $\beta$ of subhorizon
matter-density perturbations follow from the simultaneous numerical
solution of ODE \eqref{eq:ODE-DeltaM2} with boundary conditions
from \eqref{eq:DeltaM2-early} evaluated at $t_\text{start}$.
\vspace*{2mm}}%
\renewcommand{\tabcolsep}{0.4pc}  
\renewcommand{\arraystretch}{1.2} 
\begin{tabular}{c|cccccc}
\hline
$t_\text{start}$ &
$t_{p}$  &
$\beta(t_{p})$ &
$s(t_{p})$ &
$t_{p}\,h(t_{p})$ &
$\overline{w}_{X}(t_{p})$ &
$z_\text{inflect}(t_{i},t_{p})$\\[1mm]
\hline
$10^{-3}$ &
$1.37354$ &  $ 0.621646$ & $ 0.725876$ & $ 0.916630$ &
$ -0.662261$ & $0.522508$\\
$10^{-4}$ &
$1.37354$ &  $ 0.621646$ & $ 0.725876$ & $ 0.916630$ &
$ -0.662258$ & $0.522527$\\
$10^{-5}$ &
$1.37354$ &  $ 0.621646$ & $ 0.725876$ & $ 0.916630$ &
$ -0.662283$ & $0.522576$\\
$10^{-6}$ &
$1.37354$ &  $ 0.621646$ & $ 0.725876$ & $ 0.916630$ &
$ -0.662271$ & $0.522511$\\
\hline
\end{tabular}
\vspace*{2mm}
\label{tab-tmin-results}
\end{table}
\vfill
\begin{table}[h]
\caption{Numerical modified-gravity results from Fig.~\ref{fig:1}
for the Hubble
parameter $H(z)$ relative to its present value $H(0)$ and the gravity
estimator $E_{G}(z)$ defined by \eqref{eq:E_G-theo},
taking $\overline{\omega}_{M2}(t_{p})=0.25$ for the $E_{G}(z)$ values
quoted. The $E_{G}$ value for redshift $z=10^2$ is essentially
equal to the analytic result \eqref{eq:E_G-theo_early}.
For comparison, also values are given for the spatially flat
$\Lambda\text{CDM}$ model with $\Omega_{M2}(\tau_{p})=0.25$
and $E_{G}(z)$ defined by \eqref{eq:E_G-theo-LCDM}.\vspace*{2mm}}%
\renewcommand{\tabcolsep}{0.4pc}    
\renewcommand{\arraystretch}{1.2} 
\begin{tabular}{c|cccc}
\hline
$z$ & $H(z)/H(0)\,\big|_{\text{mod-grav}}$
    & $H(z)/H(0)\,\big|_{\Lambda\text{CDM}}$
    & $E_{G}^\text{\,theo}\,\big|_{\text{mod-grav}}$
    & $E_{G}^\text{\,theo}\,\big|_{\Lambda\text{CDM}}$\\[1mm]
\hline
$0$   & $1.00$ & $1.00$ & $ 0.554$ &  $ 0.541$\\
$0.25$& $1.20$ & $1.11$ & $ 0.456$ &  $ 0.418$\\
$0.5$ & $1.43$ & $1.26$ & $ 0.399$ &  $ 0.355$\\
$1$   & $1.94$ & $1.66$ & $ 0.335$ &  $ 0.298$\\
$1.5$ & $2.51$ & $2.16$ & $ 0.301$ &  $ 0.275$\\
$2$   & $3.15$ & $2.74$ & $ 0.281$ &  $ 0.265$\\
$2.5$ & $3.83$ & $3.39$ & $ 0.267$ &  $ 0.259$\\
$3$   & $4.57$ & $4.09$ & $ 0.257$ &  $ 0.256$\\
$5$   & $7.94$ & $7.40$ & $0.237$  &  $0.252$\\
$10$  & $18.9$ & $18.3$ & $0.222$  &  $0.250$\\
$10^2$& $508$  & $508$ & $0.211$  &  $ 0.250$\\
\hline
\end{tabular}
\label{tab-H-EG-results}
%
%
\end{table}

\end{document}